\documentclass[12pt, onecolumn]{IEEEtran}
\usepackage{amsmath}
\usepackage{amssymb}
\usepackage{mathrsfs}
\usepackage{cite}
\usepackage{epsfig}
\usepackage{epsf}
\usepackage{theorem}
\usepackage{graphics}
\usepackage[active]{srcltx}

\renewcommand{\baselinestretch}{1.91}

\begin{document}

\title{Adaptive Demodulation in Differentially Coherent Phase
Systems: Design and Performance Analysis
\thanks{- The first author completed his Ph.D. in ECE department at University of Toronto. This work was supported in part by a postgraduate scholarship from
the Natural Sciences and Engineering Research Council of Canada
(NSERC), an Industry Canada Fessenden postgraduate scholarship, and
in part by an Ontario Research Fund (ORF) project entitled
``Self-Powered Sensor Networks''.}
\thanks{- The material in this paper was presented
in part in the $18^{th}$ Annual IEEE Int. Symposium on Personal,
Indoor and Mobile Radio Communications (PIMRC'07), 2007
\cite{David_PIMRC2007}.}}

\author{\small \textbf{J. David Brown$^{\dag}$, Jamshid Abouei$^{\dag \dag}$, \emph{Member, IEEE}, Konstantinos N. Plataniotis$^{\dag \dag}$, \emph{Senior Member, IEEE} and
Subbarayan Pasupathy$^{\dag \dag}$, \emph{Life Fellow, IEEE}} \\
$^{\dag}$ Ottawa, Canada, Tel: 613-236-1051, Email:
david\_jw\_brown@yahoo.com\\
$^{\dag \dag}$ \small The Edward S. Rogers Sr. Dept. of Electrical and Computer Engineering, \\
University of Toronto, Toronto, Canada, Emails: \{abouei, kostas,
pas\}@comm.utoronto.ca}

\maketitle

\markboth{Submitted to IEEE Transactions on Communications, June
2010}{}

\begin{abstract}
Adaptive Demodulation (ADM) is a newly proposed rate-adaptive system
which operates without requiring Channel State Information (CSI) at
the transmitter (unlike adaptive modulation) by using adaptive
decision region boundaries at the receiver and encoding the data
with a rateless code. This paper addresses the design and
performance of an ADM scheme for two common differentially coherent
schemes: M-DPSK (M-ary Differential Phase Shift Keying) and M-DAPSK
(M-ary Differential Amplitude and Phase Shift Keying) operating over
AWGN and Rayleigh fading channels. The optimal method for
determining the most reliable bits for a given differential
detection scheme is presented. In addition, simple (near-optimal)
implementations are provided for recovering the most reliable bits
from a received pair of differentially encoded symbols for systems
using 16-DPSK and 16-DAPSK. The new receivers offer the advantages
of a rate-adaptive system, without requiring CSI at the transmitter
and a coherent phase reference at the receiver. Bit error analysis
for the ADM system in both cases is presented along with numerical
results of the spectral efficiency for the rate-adaptive systems
operating over a Rayleigh fading channel.
\end{abstract}

\begin{center}
 \vskip .3cm

\end{center}

\section{Introduction}
Rate-adaptive solutions such as adaptive modulation and incremental
redundancy have proven to be very effective at increasing the
spectral efficiency of wireless systems operating over a variety of
channels \cite{Blogh_Book2002, NandaICM0100, Zhang_ICCT2003}. These
techniques allow mobile communications systems to operate
efficiently over a wide range of channel parameters while
maintaining a target Bit Error Rate (BER), as opposed to a
non-rate-adaptive system which is designed for a worst-case
``fixed-rate'' channel. In a traditional implementation of adaptive
modulation \cite{Goldsmith_ITC1097}, the transmitter dynamically
adjusts the current level of modulation based on feedback of the
observed Channel State Information (CSI) from the receiver. The
requirement of CSI at the transmitter is one of the major
impediments of adaptive modulation systems, especially for a
high-speed mobile receiver for which the channel information changes
rapidly. Incremental redundancy \cite{Metzner_ITC0684} systems do
not require CSI at the transmitter, but instead encode data packets
using a low-rate ``mother code'' and break the message into smaller
sub-packets, sending them one-by-one until the receiver acquires
enough packets to successfully decode the message. These systems
require the receiver to decode after each sub-packet is received to
check for successful transmission and also require a buffer large
enough (in poor channel conditions) to accommodate the entire
collection of sub-packets, meaning that the complexity of the
decoding operation varies with the quality of the channel.

Adaptive Demodulation (ADM), first proposed in \cite{DavidITC0906}
and with the expanded framework in \cite{David_Thesis2008}, offers a
new rate-adaptive solution that avoids the need for CSI at the
transmitter (unlike adaptive modulation), has a fixed buffer size
(roughly equal to the size of the message), and requires the
receiver to perform a decoding operation only once (unlike
incremental redundancy). In an ADM system, the transmitter sends
data at a fixed-rate using a standard constellation and the receiver
demodulates data at a non-fixed-rate using appropriately designed
sets of decision regions chosen based on the CSI observed by the
receiver. Before transmission, each $k$-bit message is encoded using
a rateless erasure code (e.g. the Luby Transform (LT) code
\cite{LubyFOCS2002} or the Raptor code \cite{Shokrollahi_ITIT0606}).
To recover any $k$-bit message, the receiver simply buffers the
demodulated (non-erased) bits until it has accumulated
$(1+\varepsilon)k$ ``reliable'' bits (where $\varepsilon$ is a small
fixed quantity) at which point the original message can be decoded,
regardless of the erasure pattern introduced in the message. In
\cite{DavidITC0906}, the ADM-based receiver uses sets of specially
designed decision regions to demodulate some or all of the bits in a
transmitted symbol (depending upon the desired level of
reliability), and assumes a ``coherent phase reference'' is
available at the receiver. The optimal constellations and mappings
for the ADM system over an additive white Gaussian noise (AWGN)
channel and its performance for AWGN were derived in
\cite{DavidITC0906}. Particular instances of the ADM solution are
the application studies considered in \cite{Liu_ITWC0609} and
\cite{Illanko_VTC2009} for Gaussian relay channels.

In this paper, we consider how to implement the ADM system using
differentially coherent detection: an attractive alternative when a
reliable phase reference is not available. This is certainly a valid
and interesting problem when the channel changes rapidly making
phase coherence difficult or expensive to achieve. There are several
works in the literature that study the performance analysis of
differentially coherent detection in wireless RF applications (e.g.,
\cite{Stojanovic_JSAC0905}) with focus on mobile satellite
communications \cite{Lee_IT0781}, and optical communication systems
\cite{Kahn2006}. Of interest is to utilize the Differential
Phase-Shift Keying (DPSK) in mobile communications systems which
commonly circumvents the ambiguity in the phase recovery
\cite{Proakis2001}.

To the best of our knowledge, this is the first paper addressing
this rate-adaptive technique with focus on differentially coherent
detection in time-varying wireless systems. This paper studies the
design and performance analysis of a hard-decision version of the
ADM scheme for two common differentially coherent schemes: M-DPSK
and M-DAPSK (M-ary Differential Amplitude and Phase Shift Keying)
operating over AWGN and Rayleigh fading channels. The paper makes
the following contributions:

\begin{itemize}
\item [$\bullet$] The optimal method for determining the most reliable bits for a
given differential detection scheme is presented.
\item[$\bullet$] Simple (near-optimal) implementations are provided for recovering the most
reliable bits from a received pair of differentially encoded symbols
for systems using 16-DPSK and 16-DAPSK. The new receivers offer the
advantages of a rate-adaptive system, without requiring CSI at the
transmitter and a coherent phase reference at the receivers.
\item[$\bullet$] Bit error analysis for the ADM system in both cases is
presented along with numerical results of the spectral efficiency
for the rate-adaptive systems operating over a Rayleigh fading
channel.
\end{itemize}

The rest of the paper is organized as follows. Section
\ref{System_model} provides an overview of the system under
consideration and briefly introduces some ADM concepts and
terminology. Section \ref{Optimal Decision_DPSK} presents the
optimal method for determining the most reliable bits for a given
differential detection scheme in the ADM-based system with 16-DPSK.
With a similar argument as for the DPSK-ADM system model, Section
\ref{Optimal Decision_DAPSK} deals with a simple (near-optimal)
implementation for recovering the most reliable 1, 2, 3, or 4 bits
from a received pair of differentially encoded symbols for systems
using 16-DAPSK. Sections \ref{Optimal Decision_DPSK} and
\ref{Optimal Decision_DAPSK} also include the Log-Likelihood Ratio
(LLR) computation of each bit for every pair of received symbols.
Section \ref{Performance} includes the probability of error analysis
for the near-optimal detection schemes and presents the spectral
efficiency of ADM using differential detection. Finally, Section
\ref{conclusion} provides a brief summary and conclusions.


\section{System Model and Assumptions}\label{System_model}
\begin{figure} [t]
  \centering
  \includegraphics [width=5in] {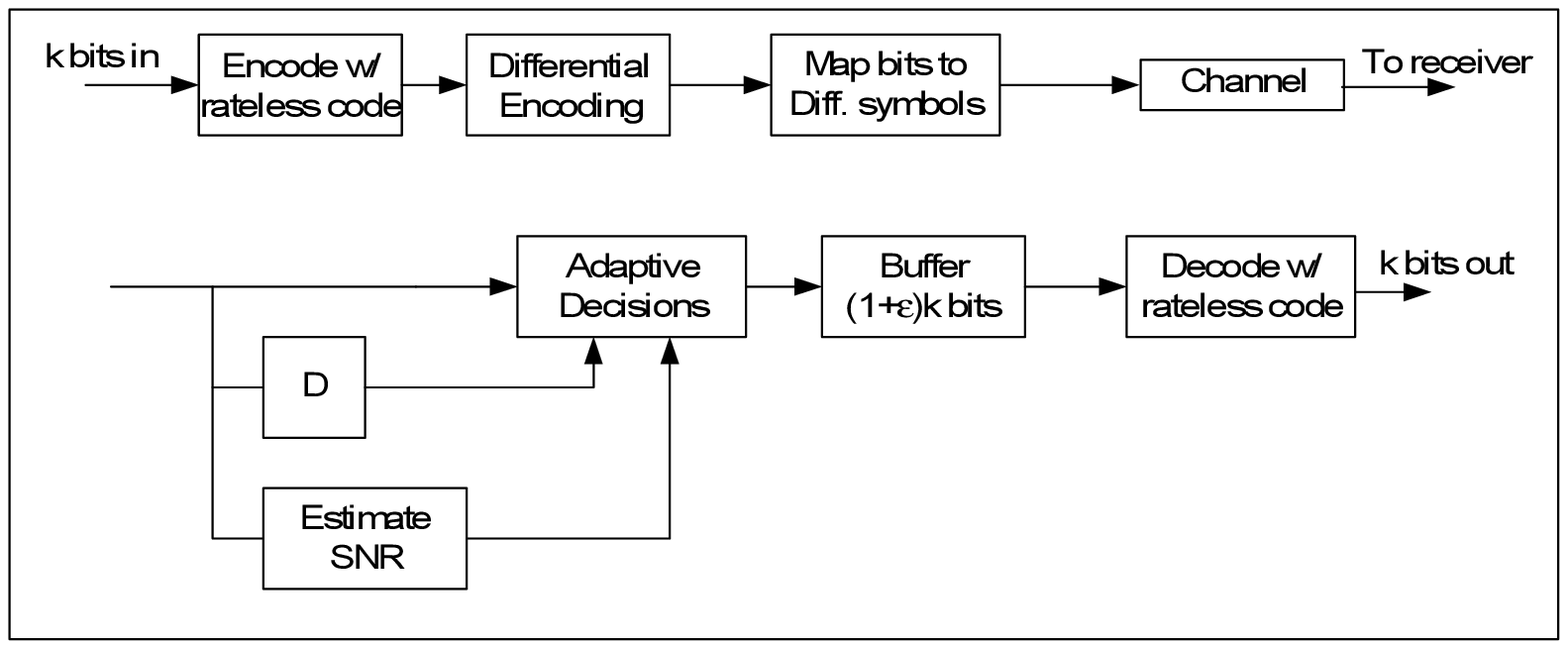}
  \caption{Block diagram of a differential ADM transmitter/receiver.}
  \label{fig: ADM_TX_RX}
\end{figure}

In this work, we consider an ADM-based single-hop wireless system
depicted in Fig. \ref{fig: ADM_TX_RX}, where the transmitter desires
to send data packets of equal length toward the corresponding
receiver using a predetermined rateless code and differential
encoding. To send a $k$-bit data packet, the transmitter first
encodes it using an LT (Luby Transform) code to produce a stream of
coded bits. Such codes are a class of sparse-graph ``rateless''
erasure codes which are fully explained in \cite{LubyFOCS2002} and
\cite{Shokrollahi_ITIT0606}. It is shown in \cite{Jamshid_BSC2010}
that LT codes offer strong benefits in using in practical wireless
networking applications with dynamic distance and position nodes. LT
codes have a very simple encoding mechanism that allows the
transmitter to produce a potentially limitless sequence of encoded
bits from the original $k$ message bits. Once the receiver
successfully recovers any $(1+\varepsilon)k$ reliable encoded
bits--where $\varepsilon$ is determined by the reliability of the
recovered bits--a message passing decoder using the
\emph{sum-product algorithm} \cite{KschischangITIT0201}, can be used
to recover the original $k$-bit message.

\begin{figure} [t]
  \centering
  \includegraphics [width=4.5in] {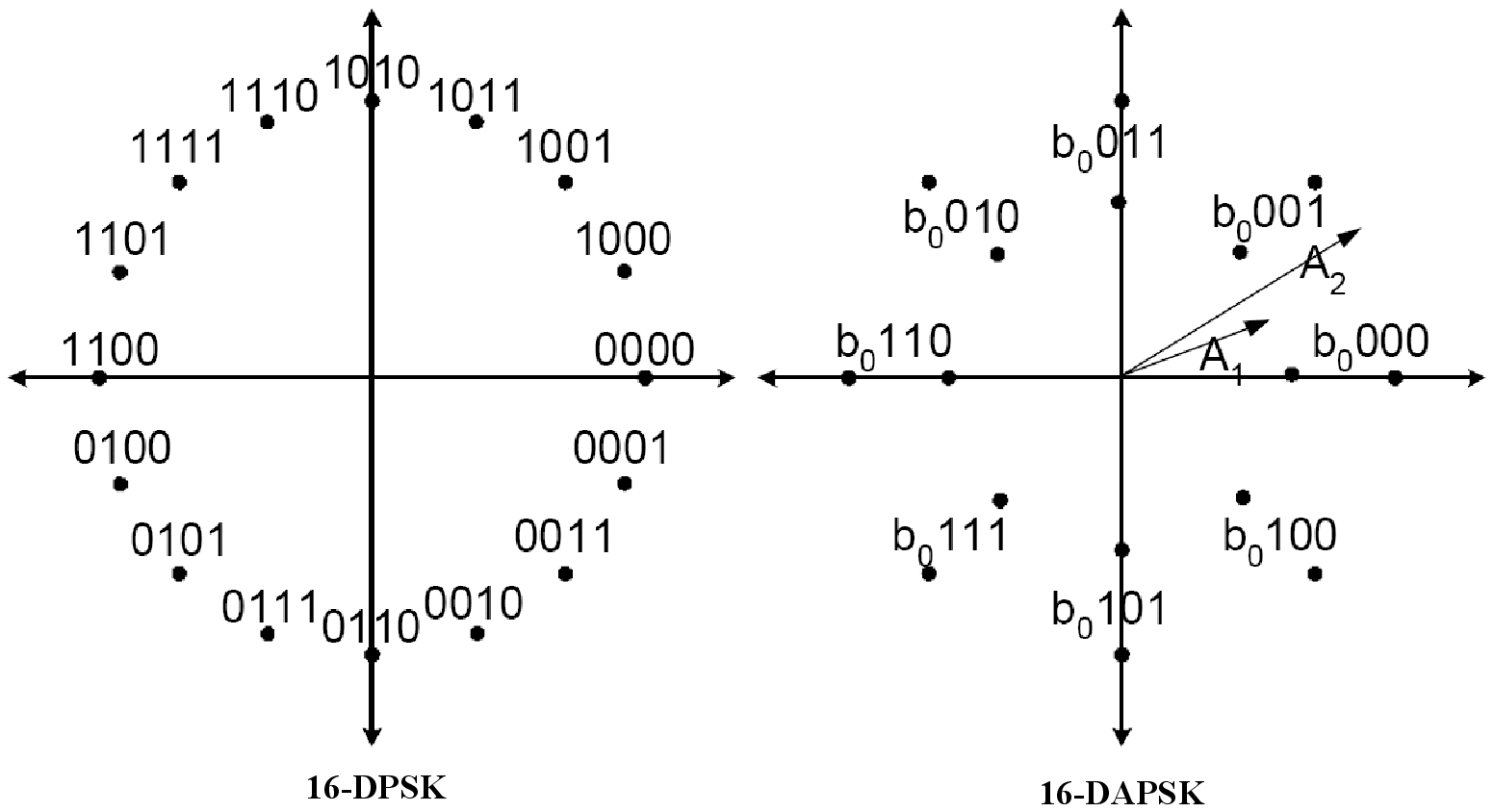}
  \caption{Mappings of differential information for 16-DPSK and 16-DAPSK.}
  \label{fig: DPSK_DAPSK}
\end{figure}

The coded bits are used to produce differentially encoded symbols.
In this paper, we consider two common differentially coherent
schemes: 16-DPSK and 16-DAPSK. For 16-DPSK, the phase difference
between successive symbols conveys 4 bits of information (here
denoted by $b_0b_1b_2b_3$) in the standard fashion of
\cite{Lindsey_Book1973} with the mapping shown in Fig. \ref{fig:
DPSK_DAPSK}, where $b_i \in \{0,1 \}$, $i=0,1,2,3$. Differential
Amplitude Phase Shift Keying is a popular alternative to DPSK and
has a performance advantage over DPSK at the expense of slightly
increased constellation complexity \cite{Xiao_JCN2006}. 16-DAPSK is
implemented as described in \cite{Xiao_JCN2006} where the amplitude
difference between successive symbols conveys 1 bit of information
(here denoted by $b_0$) and the phase difference conveys an
additional 3 bits ($b_1b_2b_3$), according to the mapping in Fig.
\ref{fig: DPSK_DAPSK}. By convention, a change in amplitude (i.e.,
consecutive symbols are transmitted on different rings) represents a
``1'' bit, while no change in amplitude represents a ``0'' bit. The
differential symbols are transmitted over an AWGN
channel\footnote{The Rayleigh fading case will be considered later
in Section \ref{Performance}. For the fading channel model and as
per the standard assumption in differential schemes, it is assumed
that the complex channel gain is constant over at least two symbol
intervals.}. For the remainder of the paper, we observe the
following notation.

\textbf{DPSK:} After matched filtering, a sample of a 16-DPSK symbol
at the receiver can be written as
\begin{equation}
y_k=s_ke^{j\theta_k}+n_k,
\end{equation}
where $s_k \triangleq e^{j\phi_k}$ represents a transmitted symbol
with phase $\phi_k$ and normalized symbol energy, $\theta_k$
represents the arbitrary (and unknown) phase introduced by the
channel, and $n_k$ is a zero-mean circularly-symmetric complex
Gaussian random variable with variance $\sigma^2 = \frac{N_0}{2}$ in
each dimension. It is assumed that $\theta_k$ is constant over at
least two symbol intervals such that $\theta_k = \theta_{k-1}$. In
addition, we denote the phase difference between consecutive 16-DPSK
symbols $s_k$ and $s_{k-1}$ by
\begin{equation}
\Delta \phi_k
=\phi_k+\theta_k-(\phi_{k-1}+\theta_{k-1})=\phi_k-\phi_{k-1}.
\end{equation}

\textbf{DAPSK:} Similarly, adopting a notation akin to
\cite{Xiao_JCN2006}, we denote two consecutive received 16-DAPSK
symbols (the $(k - 1)^{\textrm{th}}$ and $k^{\textrm{th}}$ symbols)
as follows:
\begin{eqnarray}
\label{DAPSK01}z_{k-1}&=&d_{k-1}e^{j\theta_k}+n_{k-1},\\
\label{DAPSK02}z_k&=&\alpha_k d_{k-1}e^{j\theta_k}+n_{k},
\end{eqnarray}
where $d_{k-1}$ is a complex quantity representing the magnitude and
the phase of the $(k - 1)^{\textrm{th}}$ symbol, $\alpha_k$ is a
complex quantity representing the change in magnitude and phase from
the $(k - 1)^{\textrm{th}}$ to the $k^{\textrm{th}}$ symbol, and
$n_k$ and $n_{k-1}$ are independent Gaussian noise samples as above.
We note that $|d_{k-1}|$ can take on one of only two values: $A_1
\triangleq \sqrt{\frac{2}{1 + R^2}}$ or $A_2 \triangleq
R\sqrt{\frac{2}{1 + R^2}}$, where $R\triangleq \frac{A_2}{A_1}$ is
defined as the \emph{ring ratio} in 16-DAPSK. The symbol energy is
assumed to be normalized such that $\frac{A_{1}^{2} +A_{2}^{2}}{2} =
1$. The change in magnitude, $|\alpha_k|$, can be one of three
values: $1/R$, 1, or $R$. Clearly, only certain combinations of
$|d_{k-1}|$ and $|\alpha_k|$ are allowable; for example, it would
not be possible to have $|d_{k-1}| = A_2$ and $|\alpha_k| = R$,
since this would require the $k^{\textrm{th}}$ symbol to have
magnitude not equal to either $A_1$ or $A_2$. The following ordered
pairs ($|d_{k-1}|$, $|\alpha_k|$) specify the only allowable
combinations of $|d_{k-1}|$ and $|\alpha_k|$: $(A_1,1)$, $(A_2,1)$,
$(A_1,R)$, and $(A_2,1/R)$.

\textbf{Decision Scheme:} In the ADM system with differentially
coherent detection, the receiver utilizes a $\beta$-Decision Scheme
($\beta$-DS) to recover the $\beta$ most likely bits\footnote{For
instance, for the 16-DPSK or 16-DAPSK, $\beta \in \{1,2,3,4 \}$.}
transmitted based on the observation of two consecutive
differentially encoded symbols as will be described in the
subsequent sections. Indeed, the receiver uses the instantaneous
Signal-to-Noise Ratio (SNR) to determine which $\beta$-DS to use to
demodulate a pair of symbols in order to maintain a given BER,
$P_b$, as a standard of reliability. For the above scheme,
non-demodulated (unreliable) bits are erased. Demodulated (reliable)
bits are buffered until the receiver has collected sufficient bits
for decoding at which point the decoder returns the decoded $k$-bit
message.

\section{Optimal $\beta$-Decision Schemes for 16-DPSK}\label{Optimal Decision_DPSK}
For a particular pair of differentially encoded symbols, the $\beta$
most reliable bits in the output of the $\beta$-DS should be the
$\beta$ bits with the largest magnitude Log Likelihood Ratios
(LLRs). Thus, the ``optimal'' $\beta$-DS is a device that computes
the LLRs of each bit for every pair of received symbols, compares
their magnitudes, and chooses the $\beta$ bits with the largest
LLRs. We begin with this construction and introduce several
approximations that produce simple, near-optimal $\beta$-DSs with
easy implementations that \emph{do not} require any LLR
computations.

Let $B_{i,j}$ denote the set of all differential angles $\Delta
\phi$ such that bit $b_i=j$, where $j \in \{0,1\}$. Thus, the
likelihood ratio for $b_i$ for two consecutive received symbols
$y_k$ and $y_{k-1}$ can be written as
\begin{equation}\label{LLR01}
\Lambda_{b_i}(y_k,y_{k-1})=\dfrac{\textrm{Pr}(y_k,y_{k-1}|b_i=0)}{\textrm{Pr}(y_k,y_{k-1}|b_i=1)}=\dfrac{\sum_{\Delta
\phi_m \in
B_{i,0}}\textrm{Pr}(y_k,y_{k-1}|\Delta\phi_m)}{\sum_{\Delta
\phi_{\ell} \in B_{i,1}}\textrm{Pr}(y_k,y_{k-1}|\Delta\phi_{\ell})},
\end{equation}
where $\textrm{Pr}(y_k,y_{k-1}|\Delta \phi_m)$ denotes the \emph{a
posteriori} probability of observing the pair of received symbols
$y_k$ and $y_{k-1}$ given that differential angle $\Delta \phi_m$
was encoded at the transmitter. Using a modified form of eqn. $(9)$
in \cite{Divsalar_ITC0390}, it can be shown that
\begin{equation}\label{Bessel01}
\textrm{Pr}(y_k,y_{k-1}|\Delta
\phi_m)=\dfrac{e^{-\frac{2+|y_k|^2+|y_{k-1}|^2}{2\sigma^2}}}{(2\pi
\sigma^2)^2}I_0\left(\dfrac{|y_k+y_{k-1}e^{j\Delta
\phi_m}|}{\sigma^2} \right),
\end{equation}
where $I_0(x)$ is the zeroth order modified Bessel function of the
first kind. It is revealed from (\ref{Bessel01}) that only the
Bessel term of $\textrm{Pr}(y_k,y_{k-1}|\Delta \phi_m)$ is a
function of $\Delta \phi_m$, thus allowing us to cancel the
exponential leading terms that would appear in an expansion of
(\ref{LLR01}). This results in $\Lambda_{b_i}$ being the ratio of
sums of Bessel terms, i.e.,
\begin{equation}\label{LLR02}
\Lambda_{b_i}(y_k,y_{k-1})=\dfrac{\sum_{\Delta \phi_m \in
B_{i,0}}I_0\left(\dfrac{|y_k+y_{k-1}e^{j\Delta \phi_m}|}{\sigma^2}
\right)}{\sum_{\Delta \phi_{\ell} \in
B_{i,1}}I_0\left(\dfrac{|y_k+y_{k-1}e^{j\Delta
\phi_{\ell}}|}{\sigma^2} \right)}.
\end{equation}
Note that $I_0(x)$ is a rapidly increasing function of $x$, when $x$
grows, and is approximated by $I_0(x)\approx \frac{e^x}{\sqrt{2\pi
x}}$. Thus, for moderate to large values of SNR (i.e. when
$1/\sigma^2$ is large), the sums in (\ref{LLR02}) can both be well
approximated by a single dominant term, since the argument of the
Bessel function will be in the steep part of the curve. To find
these dominant terms in (\ref{LLR02}), we observe that
\begin{equation}\label{Dominant01}
|y_k+y_{k-1}e^{j\Delta
\phi_m}|=\sqrt{|y_k|^2+|y_{k-1}|^2+2|y_k||y_{k-1}|\textrm{cos}(\phi_k-\Delta
\phi_m)},
\end{equation}
where $\phi_k$ denotes the phase difference between $y_k$ and
$y_{k-1}$. Clearly, (\ref{Dominant01}) is maximized when the
difference $|\phi_k -\Delta \phi_m|$ is minimized; thus the dominant
term in the sums of Bessel functions will be the term with encoded
differential phase $\Delta \phi_m \in B_{i,0}$ closest to the phase
difference $\phi_k$. Let $\Delta \phi_{i,0}$ denote the particular
differential phase $\Delta \phi_m \in B_{i,0}$ such that $|\phi_k
-\Delta \phi_m|$ is minimized. Similarly, let $\Delta \phi_{i,1}$
denote the particular differential phase $\Delta \phi_{\ell} \in
B_{i,1}$ such that $|\phi_k -\Delta \phi_{\ell}|$ is minimized. With
these definitions and under the high SNR approximation,
(\ref{LLR02}) can be written as
\begin{equation}\label{LLR03}
\Lambda_{b_i}(y_k,y_{k-1})\approx \dfrac{I_0
\left(\frac{1}{\sigma^2}\sqrt{|y_k|^2+|y_{k-1}|^2+2|y_k||y_{k-1}|\textrm{cos}(\phi_k-\Delta
\phi_{i,0})} \right)}{I_0
\left(\frac{1}{\sigma^2}\sqrt{|y_k|^2+|y_{k-1}|^2+2|y_k||y_{k-1}|\textrm{cos}(\phi_k-\Delta
\phi_{i,1})} \right)}.
\end{equation}

To find the $\beta$-DSs, the values of $|\ln(\Lambda_{b_i}(y_k,
y_{k-1}))|$ must be compared for all $b_i$ over the range $0 <
\phi_k \leq 2 \pi$. To simplify this comparison, we define $\Delta
\phi_{i,min}$ to be the value of $\Delta \phi_{i,0}$ or $\Delta
\phi_{i,1}$ for which $|\phi_k - \Delta \phi_{i,min}|$ is minimized
and $\Delta \phi_{i,max}$ to be the value of $\Delta \phi_{i,0}$ or
$\Delta \phi_{i,1}$ for which $|\phi_k - \Delta \phi_{i,max}|$ is
maximized. For example, if $0 < \phi_k < \pi/16$ then $\Delta
\phi_{2,min} = \Delta \phi_{2,0} = 0$ and $\Delta \phi_{2,max} =
\Delta \phi_{2,1} = -\pi/4$.  Some thought reveals that \textit{all}
$\Delta \phi_{i,min}$ will be equal for all $b_i$ for any given
value of $\phi_k$ (i.e., $\Delta \phi_{0,min} = \Delta \phi_{1,min}
= \Delta \phi_{2,min} = \Delta \phi_{3,min}$). However, this is not
true in general for $\Delta \phi_{i,max}$\footnote{Continuing the
example of $0 < \phi_k < \pi/16$, we will have $\Delta \phi_{i,min}
= 0$ for all $i$, and $\Delta \phi_{0,max} = \pi/8$, $\Delta
\phi_{1,max} = -\pi/2$, $\Delta \phi_{2,max} = -\pi/4$ (as above),
and $\Delta \phi_{3,max} = -\pi/8$.}.

Since
\begin{eqnarray} \nonumber
I_0 \left( \frac{1}{\sigma^2} \sqrt{|y_k|^2 + |y_{k-1}|^2 +
2|y_k||y_{k-1}|\cos\left( \phi_k - \Delta \phi_{i,min} \right)}
\right) ~~ > \\
~~~~~~~~~~~~~ I_0 \left( \frac{1}{\sigma^2} \sqrt{|y_k|^2 +
|y_{k-1}|^2 + 2|y_k||y_{k-1}|\cos\left( \phi_k - \Delta \phi_{i,max}
\right)} \right)~~~,
\end{eqnarray}
then we can write
\begin{eqnarray} \nonumber
|\ln(\Lambda_{b_i}(y_k, y_{k-1}))| &\approx& \ln\left(I_0 \left(
\frac{1}{\sigma^2} \sqrt{|y_k|^2 + |y_{k-1}|^2 +
2|y_k||y_{k-1}|\cos\left( \phi_k - \Delta \phi_{i,min} \right)}
\right)\right)-\\ &&  \ln\left(I_0 \left( \frac{1}{\sigma^2}
\sqrt{|y_k|^2 + |y_{k-1}|^2 + 2|y_k||y_{k-1}|\cos\left( \phi_k -
\Delta \phi_{i,max} \right)} \right)\right).
\end{eqnarray}
Noting the fact that all $\Delta \phi_{i,min}$ are equal for all
$i$, comparing the values of $|\ln(\Lambda_{b_i}(y_k, y_{k-1}))|$
can be achieved by comparing $\ln\left(I_0 \left( \frac{1}{\sigma^2}
\sqrt{|y_k|^2 + |y_{k-1}|^2 + 2|y_k||y_{k-1}|\cos\left( \phi_k -
\Delta \phi_{i,max} \right)} \right)\right)$ for all $i$. Finally,
using the fact that $\ln(x)$ and $I_0(x)$ are both monotonic
increasing, it can be shown that this comparison reduces to the
following simple decision rules:
\begin{eqnarray}
\label{eqn:DPSK_decision1}\left| \phi_k - \Delta \phi_{i,max}
\right| &>& \left| \phi_k - \Delta \phi_{j,max} \right| \rightarrow
\left| \ln\left(\Lambda_{b_i}(y_k, y_{k-1}) \right)\right| > \left|
\ln\left(\Lambda_{b_j}(y_k,
y_{k-1}) \right)\right|,\\
\label{eqn:DPSK_decision2}\left| \phi_k - \Delta \phi_{i,max}
\right| &<& \left| \phi_k - \Delta \phi_{j,max} \right| \rightarrow
\left| \ln\left(\Lambda_{b_i}(y_k, y_{k-1}) \right)\right| <
\left|\ln\left(\Lambda_{b_j}(y_k, y_{k-1}) \right)\right|.
\end{eqnarray}

\begin{figure} [htb]
  \centering
  \includegraphics [width=4in] {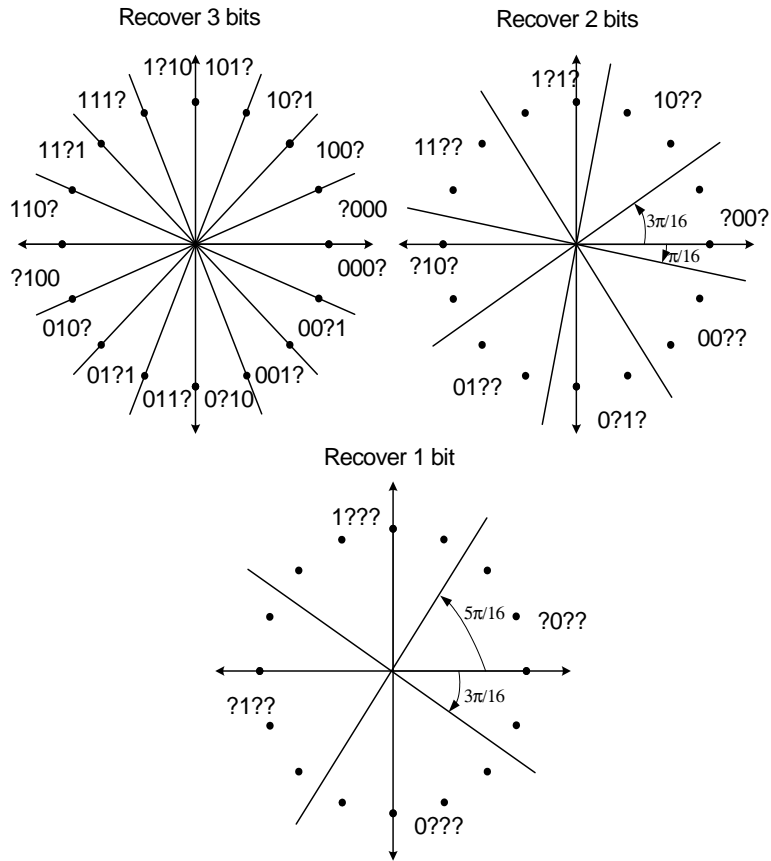}
  \caption{$\beta$-decision scheme for 16-DPSK to recover 1, 2, and 3 bits for a pair
  of differential symbols.}
  \label{16DPSK_decisions}
\end{figure}

Using the decision rules in (\ref{eqn:DPSK_decision1}) and
(\ref{eqn:DPSK_decision2}), it is a simple matter to determine the
$\beta$-DSs to recover the most likely 1, 2, and 3 bits for each
pair of 16-DPSK symbols. The receiver simply computes the phase
difference $\phi_k$ between consecutive symbols and uses this angle
to demodulate $\beta$ bits based on the regions given in
Fig.~\ref{16DPSK_decisions}, where the regions are derived from the
decision rules above.

\section{Optimal $\beta$-Decision Schemes for 16-DAPSK}\label{Optimal Decision_DAPSK}
Recalling the notation from (\ref{DAPSK01}) and (\ref{DAPSK02}), the
standard method used to demodulate 16-DAPSK is to compute the
decision statistics $r_k \triangleq \frac{|z_k|}{|z_{k-1}|}$ and
$\psi_k \triangleq \arg(z_k) - \arg(z_{k-1})$.  Typically, the $r_k$
statistic is used to make a decision on the amplitude bit ($b_0$),
while the $\psi_k$ statistic is used to decide on the phase bits
($b_1b_2b_3$). We determine the optimal $\beta$-DSs of 16-DAPSK to
find the most likely 1, 2, or 3 bits based on the observed $r_k$ and
$\psi_k$.

Proceeding in a similar fashion to Section \ref{Optimal
Decision_DPSK}, define $C_{i,j}$ to be the set of all triples
$(|d_{k-1}|,|\alpha_k|,\arg(\alpha_k))$ such that bit $b_i=j$, where
$j \in \{0,1 \}$. The likelihood ratio for $b_i$ for the received
statistics $r_k$ and $\psi_k$ is thus given by
\begin{eqnarray} \nonumber
\Lambda_{b_i}\left(r_k, \psi_k \right) & = & \frac{\textrm{Pr}(r_k,
\psi_k | b_i = 0)}{\textrm{Pr}(r_k, \psi_k | b_i = 1)}\\
\label{eqn:DAPSK_LR}& = &
\frac{\sum_{(|d_{k-1}|,|\alpha_k|,\arg(\alpha_k))_m \epsilon
C_{i,0}} \textrm{Pr}\left(r_k, \psi_k |~
(|d_{k-1}|,|\alpha_k|,\arg(\alpha_k))_m\right)}
{\sum_{(|d_{k-1}|,|\alpha_k|,\arg(\alpha_k))_{\ell} \epsilon
C_{i,1}} \textrm{Pr}\left(r_k, \psi_k |~
(|d_{k-1}|,|\alpha_k|,\arg(\alpha_k))_{\ell}\right)}.
\end{eqnarray}

Using eqn. (3.21) to (3.30) in \cite{Xiao_Thesis2004}, it can be
shown that
\begin{eqnarray} \label{eqn:DAPSK_pdf}
\textrm{Pr}\left(r_k, \psi_k |~
(|d_{k-1}|,|\alpha_k|,\arg(\alpha_k))\right) =
\frac{e^{\frac{\xi^2}{B} - P}}{(\sigma^2)^2 \pi B^3} \left( \xi^2 +
B \right) r_k,
\end{eqnarray}
where
\begin{eqnarray}
\xi^2 & \triangleq & \left(\frac{|d_{k-1}|}{\sigma^2}\right)^2
\left( 1 + |\alpha_k|^2 r_k^2 + 2|\alpha_k|r_k \cos(\psi_k -
\arg(\alpha_k)) \right), \\
P & \triangleq & \frac{|d_{k-1}|^2}{\sigma^2} (1 + |\alpha_k|^2),
\rm ~~~~~ and \\
B & \triangleq & \frac{1+r_k^2}{\sigma^2}.
\end{eqnarray}
A useful property of (\ref{eqn:DAPSK_LR}) when the \textit{a
posteriori} probabilities are given as in (\ref{eqn:DAPSK_pdf}) is
that $\Lambda_{b_i}\left(r_k, \psi_k \right)$ has the same value for
$r_k$ as for $\dfrac{1}{r_k}$.  Thus, any $\beta$-DS for 16-DAPSK
can compute $r_k' \triangleq \min(r_k, \frac{1}{r_k}) \leq 1$ and
use this value for detection without loss of accuracy.  This
simplifies the decision schemes since the analysis can be confined
to values of $r_k'$ contained within the unit circle.  Intuitively,
it makes sense that $r_k$ and $\dfrac{1}{r_k}$ would lead to the
same reliability since it should not matter the order we transition
from one ring to the other (since either way we will have one point
on the large ring and one on the small ring).

Note that $e^x$ is a rapidly increasing function of $x$, when $x$
grows. Thus, for moderate to large values of SNR and with a similar
argument as for 16-DPSK, it turns out that the sums in the numerator
and denominator of (\ref{eqn:DAPSK_LR}) can be approximated by a
single dominant term. Further simplifications arise by noting that
the denominator of (\ref{eqn:DAPSK_pdf}) and the multiplicative
$r_k$ term are common to all terms in (\ref{eqn:DAPSK_LR}) and thus
will cancel. Additionally, at high SNR we note that $\xi^2 \gg B$,
so we drop the additive $B$ term from (\ref{eqn:DAPSK_pdf}).  Thus,
the numerator and denominator of the likelihood ratio in
(\ref{eqn:DAPSK_LR}) can be approximated by appropriately chosen
dominant terms of the form $\xi^2 e^{\frac{\xi^2}{B} - P}$.  With
some thought, it is possible to identify the values of $|d_{k-1}|$,
$|\alpha_k|$, and $\arg(\alpha_k)$ that maximize
(\ref{eqn:DAPSK_pdf}) for any particular pair of $(r_k,\psi_k)$.

\subsection{Simple Estimate for Choosing the Differential Amplitude Threshold}
Before determining the $\beta$-DSs for 16-DAPSK, we develop an
interesting result obtained from the high SNR estimate of
(\ref{eqn:DAPSK_LR}). In standard 16-DAPSK systems, the decision for
the ``differential amplitude bit" $b_0$ is to choose $b_0 = 0$ when
$r_k'
> \Delta_0$ and to choose $b_0 = 1$ when $r_k' < \Delta_0$, where
$\Delta_0$ is an appropriately chosen threshold based on the ring
ratio $R$.  An optimal value of $\Delta_0$ can be found by
numerically evaluating (\ref{eqn:DAPSK_LR}) for $i=0$ where
$\Delta_0$ is the value of $r_k'$ such that
$\Lambda_{b_0}\left(r_k', \psi_k \right) = 1$, i.e., $b_0 = 0$ and
$b_0 = 1$ are equally likely. The high SNR estimate of
(\ref{eqn:DAPSK_LR}) can be used to develop a remarkably simple (and
accurate) estimate for the threshold $\Delta_0$. At high SNR (and
for the range $1/R < r_k' < 1$), the dominant terms in
(\ref{eqn:DAPSK_LR}) are $\textrm{Pr}(r_k, \psi_k |~ (A_1,1,\phi))$
in the numerator and $\textrm{Pr}(r_k, \psi_k |~ (A_2,1/R,\phi))$ in
the denominator, with $\phi$ representing the differential angle
closest to $\psi_k$. A final approximation can be made by observing
that $\phi$ will always be within $\pm \pi/8$ of differential angle
$\psi_k$; using this fact, we use $\cos(\psi_k - \phi) \approx 1$.
Thus, at high SNR we can write
\begin{eqnarray} \label{eqn:DAPSK_highSNR_b0}
\Lambda_{b_0}\left(r_k, \psi_k \right) &\approx& e^Z \frac{ \left(
\frac{A_1}{\sigma^2}\right)^2 \left( 1 + r_k^2 + 2 r_k \right)} {
\left( \frac{A_2}{\sigma^2}\right)^2 \left( 1 + r_k^2/R^2 + 2r_k/R
\right)}\\
&\stackrel{(a)}{=}&\dfrac{e^Z}{R^2}\left(\dfrac{1+r_k}{1+\frac{r_k}{R}}\right)^2,
\end{eqnarray}
where $(a)$ comes from $R\triangleq \frac{A_2}{A_1}$, and
\begin{eqnarray} \label{eqn:Zis0}
Z \triangleq
\frac{1}{\sigma^2(1+r_k^2)}\left(A_2^2\left(\frac{1}{R}-
r_k\right)^2-A^2_1(1- r_k)^2 \right).
\end{eqnarray}

To find $\Delta_0$, we want to find the $r_k$ such that the estimate
in (\ref{eqn:DAPSK_highSNR_b0}) is equal to 1.  At high SNR, the
behavior of the exponential determines, to a large extent, whether
$\Lambda_{b_0}\left(r_k, \psi_k \right)$ is greater than or less
than 1.  That is, to a reasonable approximation, $e^Z \gg 1$ when
$r_k
> \Delta_0$ and $e^Z \ll 1$ when $r_k < \Delta_0$ (at high SNR); due
to the dominant behavior of the exponential, sign$(Z)$ gives a
reasonable estimate of whether or not $\Lambda_{b_0}\left(r_k,
\psi_k \right)$ exceeds 1.  Consequently, a simple estimate for
$\Delta_0$ is found by solving for $r_k$ when $Z=0$.  Solving
(\ref{eqn:Zis0}) when $Z=0$, results in
\begin{eqnarray}
\Delta_0 \approx r_k|_{_{Z=0}} = \dfrac{2}{1+R}.
\end{eqnarray}
This approximation has been verified to be accurate to within two
percent of the optimum value obtained through numerical solution
\cite{David_Thesis2008} (over the useful range $1.5 < R < 2.5$). For
$R=2$, results in \cite{Chow_EL0892} (and numerical evaluation of
(\ref{eqn:DAPSK_LR})) report an optimal value of $\Delta_0 = 0.68$
while the estimate gives $\Delta_0 \approx 2/3 = 0.667$.

\subsection{Optimal $\beta$-Decision Scheme}
To compute the $\beta$-DSs for 16-DAPSK, we first identify the
dominant terms in the numerator and denominator of
(\ref{eqn:DAPSK_LR}). Then, we use these estimates to compare the
likelihood ratios of each bit in a similar fashion to the way
$\Delta_0$ was determined above. Adopting a notation similar to what
was used in Section \ref{Optimal Decision_DPSK}, let $\Delta
\psi_{i,0}$ denote the particular phase $\arg(\alpha_k)$ (where
$(|d_{k-1}|,|\alpha_k|,\arg(\alpha_k)) ~ \in ~ C_{i,0})$ such that
$|\psi_k - \Delta \psi_{i,0}|$ is minimized, and define $\Delta
\psi_{i,1}$ in the same manner. Furthermore, denote by $\Delta
\psi_{i,min}$ the value of $\Delta \psi_{i,0}$ or $\Delta
\psi_{i,1}$ for which $|\psi_k - \Delta \psi_{i,min}|$ is minimized,
and denote by $\Delta \psi_{i,max}$ the value of $\Delta \psi_{i,0}$
or $\Delta \psi_{i,1}$ for which $|\psi_k - \Delta \psi_{i,max}|$ is
maximized.

An example computation comparing likelihood ratios is given as
follows. For $\Delta_0 < r_k < 1$, it can be shown that for $i=1,2,$
or $3$, the numerator of (\ref{eqn:DAPSK_LR}) is dominated by
$\textrm{Pr}(r_k, \psi_k |~ (A_1,1,\Delta \psi_{i,0}))$ and the
denominator is dominated by $\textrm{Pr}(r_k, \psi_k |~
(A_1,1,\Delta \psi_{i,1}))$. Additionally, it can be shown that for
$i=0$, the numerator is dominated by $\textrm{Pr}(r_k, \psi_k |~
(A_1,1,\Delta \psi_{0,min}))$ and the denominator is dominated by
$\textrm{Pr}(r_k, \psi_k |~ (A_2,1/R,\Delta \psi_{0,min}))$.  Now
since one of $\textrm{Pr}(r_k, \psi_k |~ (A_1,1,\Delta \psi_{i,0}))$
and $\textrm{Pr}(r_k, \psi_k |~ (A_1,1,\Delta \psi_{i,1}))$
\textit{must be} equal to $\textrm{Pr}(r_k, \psi_k |~ (A_1,1,\Delta
\psi_{i,min}))$ (with the other equal to $\textrm{Pr}(r_k, \psi_k |~
(A_1,1,\Delta \psi_{i,max}))$), then comparing
$|\ln(\Lambda_{b_0}(r_k, \psi_k ))|$ with $|\ln(\Lambda_{b_i}(r_k,
\psi_k ))|$ ($i \neq 0$) amounts to evaluating the difference
\begin{eqnarray} \nonumber
& & |\ln(\Lambda_{b_i}(r_k, \psi_k ))| - |\ln(\Lambda_{b_0}(r_k,
\psi_k ))| \\ \nonumber & = & \ln\left(\textrm{Pr}(r_k, \psi_k |~
(A_1,1,\Delta \psi_{i,min})) \right) - \ln\left(\textrm{Pr}(r_k,
\psi_k |~ (A_1,1,\Delta \psi_{i,max})) \right) \\ \nonumber & & -
\left( \ln\left(\textrm{Pr}(r_k, \psi_k |~ (A_1,1,\Delta
\psi_{0,min}))\right)
- \ln\left(\textrm{Pr}(r_k, \psi_k |~ (A_2,1/R,\Delta \psi_{0,min}))\right) \right) \\
&\stackrel{(a)}{=}& \ln\left(\textrm{Pr}(r_k, \psi_k |~
(A_2,1/R,\Delta \psi_{0,min}))\right) - \ln\left(\textrm{Pr}(r_k,
\psi_k |~ (A_1,1,\Delta \psi_{i,max})) \right),
\label{eqn:DAPSK_simplified}
\end{eqnarray}
where $(a)$ comes from the fact that $\textrm{Pr}(r_k, \psi_k |~
(A_1,1,\Delta \psi_{0,min})) = \textrm{Pr}(r_k, \psi_k |~
(A_1,1,\Delta \psi_{i,min}))$, and noting that $\Delta \psi_{i,min}$
are equal for all $i ~ \in ~ \{0, 1, 2, 3 \}$. Finding the sign of
the difference in (\ref{eqn:DAPSK_simplified}) can be computed by
evaluating the quotient
\begin{eqnarray}
L \triangleq \frac{\textrm{Pr}(r_k, \psi_k |~ (A_2,1/R,\Delta
\psi_{0,min}))}{\textrm{Pr}(r_k, \psi_k |~ (A_1,1,\Delta
\psi_{i,max}))} \stackrel{(a)}{=} \frac{R^2 + r_k^2 +2r_k R
\cos(\psi_k - \psi_{0,min})}{1+r_k^2 + 2r_k \cos(\psi_k -
\psi_{i,max})}e^Z,
\end{eqnarray}
where $Z \triangleq r_k(A_1/\sigma^2)^2(2R - 2\cos(\psi_k -
\psi_{i,max}) - r_k(R^2-1))$, and $(a)$ follows from the estimate
$\cos(\psi_k - \psi_{i,min}) \approx 1$. Using the same arguments
made when evaluating the quotient in (\ref{eqn:DAPSK_highSNR_b0}),
at high SNR sign$(Z)$ gives a reasonable estimate of whether or not
$L > 1$. This results in the following decision rule (when $\Delta_0
< r_k < 1$):  if $r_k < \frac{2}{R^2-1}(R-\cos(\psi_k -
\psi_{i,max}))$ then $b_i$ is more likely than $b_0$; otherwise,
$b_0$ is more likely.

Using this technique to compare other likelihood ratios over other
ranges of $r_k$ we obtain the following simple set of decision rules
to determine, for a given $(r_k, \psi_k)$, which bits are the most
reliable:

$\bullet$ \emph{Comparing} $|\ln(\Lambda_{b_i}(r_k, \psi_k ))|$
\emph{with} $|\ln(\Lambda_{b_j}(r_k, \psi_k ))|$ \emph{for} $i,j ~
\in ~ \{1,2,3\}$, $i \neq j$:
\begin{eqnarray} \label{eqn:DAPSK_Decision1}
|\psi_k - \Delta \psi_{i,max}| > |\psi_k - \Delta \psi_{j,max}|
\rightarrow |\ln(\Lambda_{b_i}(r_k, \psi_k ))| > |\ln
(\Lambda_{b_j}(r_k, \psi_k ))|.
\end{eqnarray}

$\bullet$ \emph{Comparing} $|\ln(\Lambda_{b_0}(r_k, \psi_k ))|$
\emph{with} $|\ln(\Lambda_{b_i}(r_k, \psi_k ))|$ \emph{for} $i ~\in
~ \{1,2,3\}$:

~~~~~$i)$ \emph{When} $r_k' > \Delta_0$:
\begin{eqnarray} \label{eqn:DAPSK_Decision2}
r_k' > \frac{2\left(R-\cos\left(\psi_k - \psi_{i,max}
\right)\right)}{R^2 - 1} \rightarrow |\ln(\Lambda_{b_0}(r_k, \psi_k
))| > |\ln (\Lambda_{b_i}(r_k, \psi_k ))|.
\end{eqnarray}

~~~~~$ii)$ \emph{When} $r_k' < \Delta_0$:
\begin{eqnarray} \label{eqn:DAPSK_Decision3}
r_k' > \frac{2\left(R\cos\left(\psi_k - \psi_{i,max}
\right)-1\right)}{R^2 - 1} \rightarrow |\ln(\Lambda_{b_i}(r_k,
\psi_k ))| > |\ln (\Lambda_{b_0}(r_k, \psi_k ))|,
\end{eqnarray}
\emph{where} $\Delta_0$ \emph{is the decision threshold for} $b_0$.

The decision rules in (\ref{eqn:DAPSK_Decision1}) to
(\ref{eqn:DAPSK_Decision3}) are intuitively satisfying. When
comparing only bits 1, 2, and 3 (i.e., the bits determined by the
differential angle), rule (\ref{eqn:DAPSK_Decision1}) is identical
to the decision rule for 16-DPSK (rule (\ref{eqn:DPSK_decision1})),
as one might expect.  When $\Delta_0 < r_k' < 1$, rule
(\ref{eqn:DAPSK_Decision2}) indicates that as $(\psi_k -
\psi_{i,max})$ increases\footnote{An increase in $(\psi_k -
\psi_{i,max})$ means that the nearest differential angle that will
result in an error has become even further away from the observed
differential angle.  This indicates that the reliability of $b_i$
has increased.} (i.e., $b_i$ becomes more reliable), then the right
hand side of (\ref{eqn:DAPSK_Decision2}) will increase, meaning that
$r_k'$ must be large (i.e., close to 1) in order for $b_0$ to be
more reliable than $b_i$.  Finally, when $r_k'<\Delta_0$, rule
(\ref{eqn:DAPSK_Decision3}) indicates that as $(\psi_k -
\psi_{i,max})$ increases (i.e., $b_i$ becomes more reliable) then
the right hand side of (\ref{eqn:DAPSK_Decision3}) decreases,
meaning that $r_k'$ must decrease ($b_0$ becoming more reliable as
$r_k'$ gets further away from $\Delta_0$) in order for $b_0$ to be
more reliable than $b_i$.

Despite the relative simplicity of the rules in
(\ref{eqn:DAPSK_Decision1}) to (\ref{eqn:DAPSK_Decision3}) (compared
to calculating and comparing LLRs), it turns out that when the rules
are evaluated over all possible $(r_k,\psi_k)$ they do not produce
$\beta$-DSs that are particularly simple to implement, unlike
16-DPSK\footnote{The $\beta$-DSs so produced have non-linear
boundaries specified by the polar form equations in
(\ref{eqn:DAPSK_Decision2}) and (\ref{eqn:DAPSK_Decision3})}. In the
next subsection, we propose simple $\beta$-DSs for 16-DAPSK based on
a heuristic analysis; subsequent judicious application of the rules
in (\ref{eqn:DAPSK_Decision1}) to (\ref{eqn:DAPSK_Decision3}) allow
us to lend some mathematical rigor to the simplified decision
schemes.

\subsection{Simple $\beta$-Decision Scheme for 16-DAPSK}
Recalling from the definition of the threshold $\Delta_0$, we
observe that when $r_k'$ is close in value to $\Delta_0$, the bit
$b_0$ is very unreliable. In this case, our $\beta$-DS (for $\beta <
4$) should always drop $b_0$ as the most unreliable bit (we will
specify below precisely how close $r_k'$ needs to be to $\Delta_0$
for this to apply). Thus for $r_k'$ ``close" to $\Delta_0$,
decisions on $b_1$, $b_2$, and $b_3$ can be made using
(\ref{eqn:DAPSK_Decision1}), resulting in a simple set of
$\beta$-DSs (similar to the DPSK regions in Fig.
\ref{16DPSK_decisions}).  Conversely, $b_0$ is at its most reliable
when $r_k'$ is ``close" to 1 or $r_k'$ is very small. So for these
values of $r_k'$, the $\beta$-DS should always demodulate $b_0$ and
demodulate $\beta$-1 bits of $b_1$, $b_2$, and $b_3$, again using
(\ref{eqn:DAPSK_Decision1}). Finally, we propose a ``transition
region" that applies for $r_k'$ in some region between 1 (i.e.,
$b_0$ reliable) and $\Delta_0$ (i.e., $b_0$ unreliable) and also for
$r_k'$ in a region between $\Delta_0$ and 0 (i.e., $b_0$ very
reliable once again).  This transition region can be viewed as a
hybrid of two regions: one region where $b_0$ is always erased and
another region where $b_0$ is always kept.  The decision scheme for
the transition region is derived by careful consideration of the
behavior of the optimum decision schemes that would be obtained by
directly evaluating the likelihood ratios of (\ref{eqn:DAPSK_LR})
over all $(r_k,\psi_k)$ (or directly implementing the rules in
(\ref{eqn:DAPSK_Decision1}) to (\ref{eqn:DAPSK_Decision3})). In
actual fact, the optimum decision schemes transition gradually
between the ``$b_0$ reliable" and ``$b_0$ unreliable" regions in a
continuous manner, passing through our proposed "transition region"
for certain particular value(s) of $r_k'$. Intuitively the proposed
transition region is a satisfying ``halfway" point between $b_0$
reliable and $b_0$ unreliable.

\begin{figure} [t]
  \centering
  \includegraphics [width=4.85in] {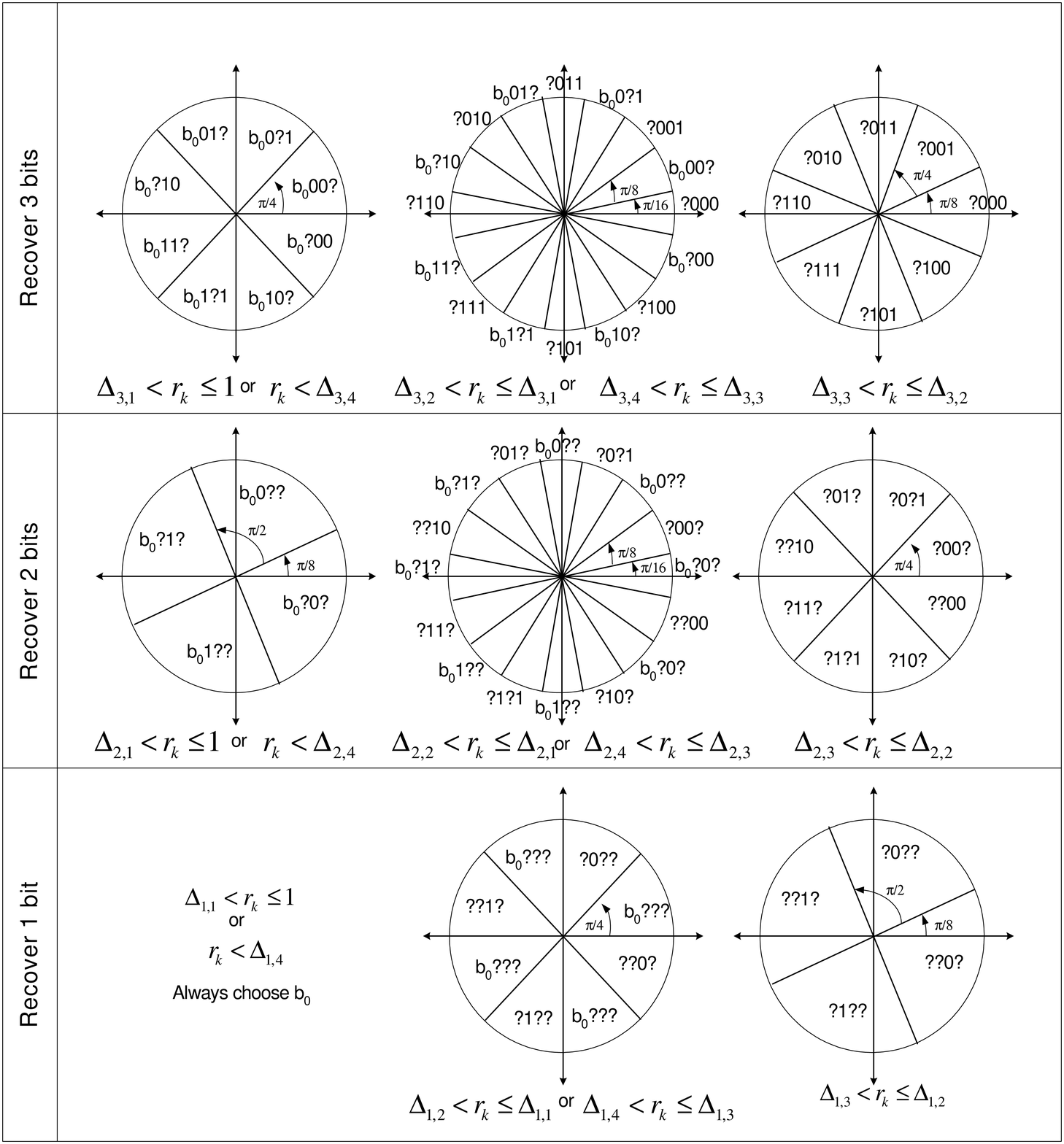}
  \caption{$\beta$-Decision Scheme for 16-DAPSK to recover 1, 2, and 3 bits for a pair
  of differential symbols.}
  \label{16DAPSK_decisions}
\end{figure}

The simplified $\beta$-DSs are given in Fig.~\ref{16DAPSK_decisions}
for $\beta~\in~\{1, 2, 3\}$, where the ``$b_0$ reliable" region
appears on the far left, the ``transition region" appears in the
middle, and the ``$b_0$ unreliable" region appears on the far right
for each $\beta$-DS.  Fig.~\ref{16DAPSK_decisions} also introduces
the threshold terms $\Delta_{\beta,1}$, $\Delta_{\beta,2}$,
$\Delta_{\beta,3}$, and $\Delta_{\beta,4}$ that are used by the
decision schemes that recover $\beta$ bits. For $\Delta_{\beta,1} <
r_k' \leq 1$ or $r_k' < \Delta_{\beta,4}$, $b_0$ is sufficiently
reliable that it is always kept, while for $\Delta_{\beta,3} < r_k'
\leq \Delta_{\beta,2}$, $b_0$ is sufficiently unreliable that it is
always discarded. Also, for all other values of $r_k'$, the
transition region is used.  A method is presented below to find
$\Delta_{\beta,1}$, $\Delta_{\beta,2}$, $\Delta_{\beta,3}$, and
$\Delta_{\beta,4}$ using (\ref{eqn:DAPSK_Decision2}) and
(\ref{eqn:DAPSK_Decision3}).

\begin{figure} [htb]
  \centering
  \includegraphics [width=5in] {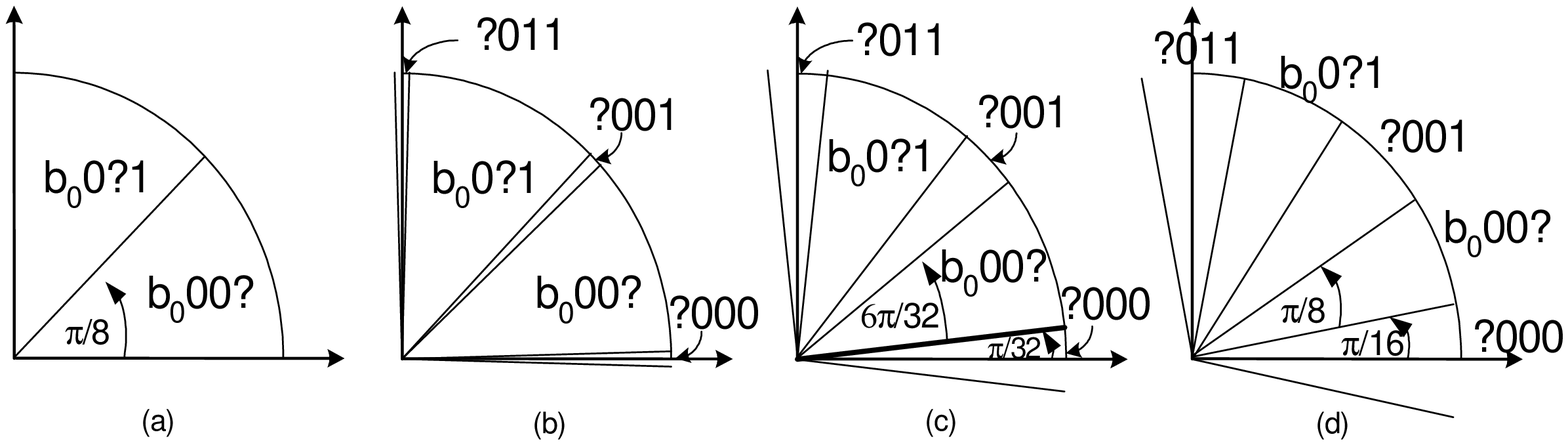}
  \caption{Progression from $b_0$ reliable to ``transition region".}
  \label{16DAPSK_transition_3DRS}
\end{figure}

\begin{figure} [htb]
  \centering
  \includegraphics [width=5in] {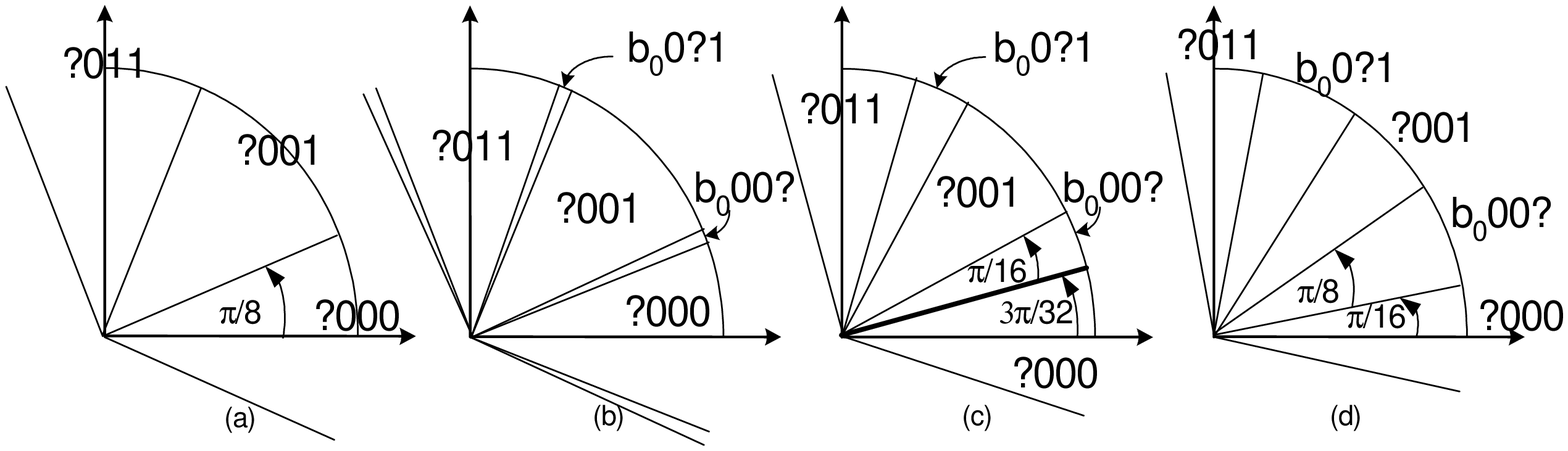}
  \caption{Progression from $b_0$ unreliable to ``transition region".}
  \label{16DAPSK_transition_3DRS_2}
\end{figure}

Consider the 3-DS shown in Fig.~\ref{16DAPSK_decisions}.  As $r_k'$
begins decreasing (from 1), an optimal 3-DS would begin to
transition from the ``$b_0$ reliable" region in a continuous fashion
to the hybrid (middle) region.  Fig.~\ref{16DAPSK_transition_3DRS}
shows this progression; Fig.~\ref{16DAPSK_transition_3DRS}(a) shows
a section of the $b_0$ reliable region, while
Fig.~\ref{16DAPSK_transition_3DRS}(b) shows the beginning of a
transition to the hybrid region.  In
Fig.~\ref{16DAPSK_transition_3DRS}(b) we see that over the angles
for which $b_3$ is most reliable (as determined from rule
(\ref{eqn:DAPSK_Decision1})), $b_0$ is dropped in favor of $b_3$
when $r_k'$ decreases by a sufficient amount.  As $r_k'$ continues
to decrease, the angle over which $b_3$ is selected (in favor of
$b_0$) grows as depicted in Fig.~\ref{16DAPSK_transition_3DRS}(c).
Eventually this angle increases to such an extent that the decision
region becomes the hybrid (middle) region shown in
Fig.~\ref{16DAPSK_transition_3DRS}(d).

For the heuristic decision schemes depicted in
Fig.~\ref{16DAPSK_decisions} we opt use the ``$b_0$ reliable" region
when the angle over which $b_3$ is selected is small (as shown in
Fig.~\ref{16DAPSK_transition_3DRS}(b)).  We switch to the hybrid
(middle) region when this angle has increased exactly halfway from
zero to $\pi/16$ (where $\pi/16$ is the angle of the $b_3$ region
for the hybrid region as shown in
Fig.~\ref{16DAPSK_transition_3DRS}(d)).  This halfway point occurs
as shown in Fig.~\ref{16DAPSK_transition_3DRS}(c), where the angle
of the $b_3$ region is $\pi/32$.  The bold line in
Fig.~\ref{16DAPSK_transition_3DRS}(c) depicts the boundary between
choosing $b_3$ and choosing $b_0$; thus, the bold line represents
the line such that $\Lambda_{b_0}(r_k, \pi/32 ) = \Lambda_{b_3}(r_k,
\pi/32 )$.  To determine the $r_k$ for which this equivalence
occurs, we can use rule (\ref{eqn:DAPSK_Decision2}) \footnote{Notice
that we use (\ref{eqn:DAPSK_Decision2}) since here we are addressing
the case where $r_k' > \Delta_0$ as we have implicity assumed that
$r_k'$ is large.} by direct substitution such that $r_k' =
\frac{2\left(R-\cos\left(\psi_k - \psi_{3,max} \right)\right)}{R^2 -
1}$, where $\psi_k = \pi/32$ and $\psi_{3,max} = \pi/4$ (the correct
value for $\psi_{3,max}$ can be determined from Fig.\ref{fig:
DPSK_DAPSK}(b)).  This value of $r_k$ is precisely the threshold
$\Delta_{3,1}$.

As another example, to find the threshold $\Delta_{3,2}$,
Fig.~\ref{16DAPSK_transition_3DRS_2} proves useful as it depicts the
transition from the ``$b_0$ unreliable" region to the hybrid region.
Once again, Fig.\ref{16DAPSK_transition_3DRS_2}(c) shows the halfway
point between these two regions such that the angle of the ``$b_0$
unreliable" region has decreased from $4\pi/32$ to $3\pi/32$:
exactly halfway to the angle $2\pi/32$ shown in
Fig.\ref{16DAPSK_transition_3DRS_2}(d).  Using rule
(\ref{eqn:DAPSK_Decision2}) we can find $r_k' =
\frac{2\left(R-\cos\left(\psi_k - \psi_{3,max} \right)\right)}{R^2 -
1}$, where $\psi_k = 3\pi/32$ and $\psi_{3,max} = \pi/4$.

In a similar fashion, the remaining thresholds for all
$\Delta_{i,j}$ can be computed. Analytical equations for these
thresholds computed in this manner are given in Table
\ref{ThresholdTable}. The equations for the thresholds are modified
such that $\Delta_{\beta,1}$ and $\Delta_{\beta,2}$ can be no larger
than $1$ and $\Delta_{\beta,3}$ and $\Delta_{\beta,4}$ must be
greater than $0$. This modification reflects the reality that $0
\leq r_k' \leq 1$. The table also evaluates the threshold equations
for a ring ratio of $R=2$ which is the optimum ring ratio for
standard 16-DAPSK in Rayleigh fading, as per \cite{Chow_EL0892}.
Observe that some of the thresholds are equal to 1 or 0 indicating
that some of the regions depicted in Fig.~\ref{16DAPSK_decisions}
are not used. Since Fig.~\ref{16DAPSK_decisions} is general and
applies for any $R$, not all regions will be used for every $R$
since the thresholds themselves are functions of $R$ by virtue of
Table \ref{ThresholdTable}.

\begin{table} [htb]
  \centering
  \caption{Equations to compute thresholds $\Delta_{i,j}$ for 16-DAPSK.}
  \begin{tabular}{c c c c c}
  \hline
  $\Delta_{i,j}$ & \vline & Equation & \vline & for $R=2$ \\
  \hline \hline
  $\Delta_{3,1}$ & \vline & $\min\left(1,\frac{2\left(R-\cos\left(\pi/32 - \pi/4 \right)\right)}{R^2 -
1}\right)$ & \vline & 0.818 \\
  $\Delta_{3,2}$ & \vline & $\min\left(1,\frac{2\left(R-\cos\left(3\pi/32 - \pi/4 \right)\right)}{R^2 -
1}\right)$ & \vline & 0.745 \\
  $\Delta_{3,3}$ & \vline & $\max\left(0,\frac{2\left(R\cos\left(3\pi/32 - \pi/4 \right)-1\right)}{R^2 -
1}\right)$ & \vline & 0.509 \\
  $\Delta_{3,4}$ & \vline & $\max\left(0,\frac{2\left(R\cos\left(\pi/32 - \pi/4 \right)-1\right)}{R^2 -
1}\right)$ & \vline & 0.364 \\
  $\Delta_{2,1}$ & \vline & $\min\left(1,\frac{2\left(R-\cos\left(3\pi/32 - (-\pi/4) \right)\right)}{R^2 -
1}\right)$ & \vline & 1 \\
  $\Delta_{2,2}$ & \vline & $\min\left(1,\frac{2\left(R-\cos\left(\pi/32 - (-\pi/4) \right)\right)}{R^2 -
1}\right)$ & \vline & 0.910 \\
  $\Delta_{2,3}$ & \vline & $\max\left(0,\frac{2\left(R\cos\left(\pi/32 - (-\pi/4) \right)-1\right)}{R^2 -
1}\right)$ & \vline & 0.179 \\
  $\Delta_{2,4}$ & \vline & $\max\left(0,\frac{2\left(R\cos\left(3\pi/32 - (-\pi/4) \right)-1\right)}{R^2 -
1}\right)$ & \vline & 0 \\
  $\Delta_{1,1}$ & \vline & $\min\left(1,\frac{2\left(R-\cos\left(3\pi/8 - (-\pi/4) \right)\right)}{R^2 -
1}\right)$ & \vline & 1 \\
  $\Delta_{1,2}$ & \vline & $\min\left(1,\frac{2\left(R-\cos\left(3\pi/16 - (-\pi/4) \right)\right)}{R^2 -
1}\right)$ & \vline & 1 \\
  $\Delta_{1,3}$ & \vline & $\max\left(0,\frac{2\left(R\cos\left(3\pi/16 - (-\pi/4) \right)-1\right)}{R^2 -
1}\right)$ & \vline & 0 \\
  $\Delta_{1,4}$ & \vline & $\max\left(0,\frac{2\left(R\cos\left(3\pi/8 - (-\pi/4) \right)-1\right)}{R^2 -
1}\right)$ & \vline & 0 \\
  \hline
  \end{tabular}
  \label{ThresholdTable}
\end{table}

As a final note, we have verified that the simplified decision
schemes of Fig.~\ref{16DAPSK_decisions} perform nearly as well as
the optimum (and complex) schemes obtained by numerically evaluating
(\ref{eqn:DAPSK_LR}) or directly implementing the rules in
(\ref{eqn:DAPSK_Decision1}) to (\ref{eqn:DAPSK_Decision3}).
Numerical results demonstrating this fact are provided in Section
\ref{Performance}.

\section{Performance of Differentially Coherent ADM}\label{Performance}
\begin{figure} [t]
  \centering
  \includegraphics [width=4.5in] {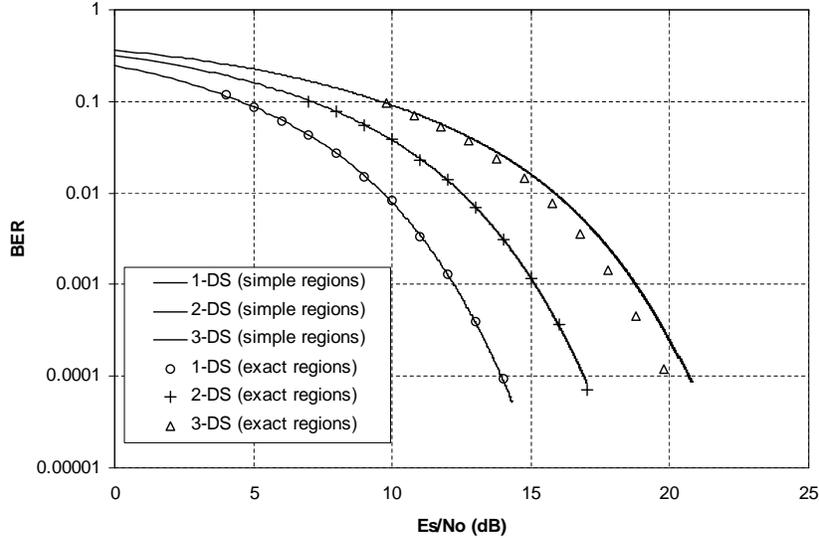}
  \caption{Performance of the optimum decision schemes for 16-DAPSK
  and ``simple" decision schemes.}
  \label{16DAPSK_estimate_vs_exact}
\end{figure}

To evaluate the performance of the ADM system for 16-DPSK and
16-DAPSK, we first examine the BER for each $\beta$ decision scheme
in AWGN.  For 16-DAPSK, the BER can be computed by numerically
integrating the density function in (\ref{eqn:DAPSK_pdf}) over the
appropriate regions of the simplified $\beta$-DSs as defined in
Fig.~\ref{16DAPSK_decisions}. Additionally, we choose $R=2$ which is
the optimum ring ratio for 16-DAPSK in Rayleigh fading
\cite{Chow_EL0892}. The effect of the ring ratio in the proposed ADM
system will be considered later.
Fig.~\ref{16DAPSK_estimate_vs_exact} illustrates the probability of
error for the simplified $\beta$ decision schemes for 16-DAPSK with
$R=2$. Also shown in Fig.~\ref{16DAPSK_estimate_vs_exact} are BER
curves (obtained through simulation) for the 16-DAPSK system using
``optimal" $\beta$ decision schemes that evaluate and compare the
exact LLRs for every pair of received symbols.  For $\beta=1$ and
$\beta=2$ we observe virtually \textit{no difference} in BER
performance between the optimal and simplified $\beta$-DSs.  For
$\beta=3$, the simplified scheme exhibits a loss of approximately
$0.6$dB at high SNR compared to the optimal scheme.  Evaluating the
BER at other values of $R$ yield similar results (negligible loss
for $\beta=1$ and $\beta=2$ and a small but measurable loss for
$\beta=3$ at high SNR).  These results suggest that the simplified
$\beta$-DSs perform nearly as well as the optimum (and complex)
schemes obtained by numerically evaluating (\ref{eqn:DAPSK_LR}).

For 16-DPSK, the probability of symbol error for a $\beta$-DS,
$P_{S,\beta}$, can be computed as
\begin{eqnarray} \label{eqn:16DPSK_integral1}
P_{S,3} & = & 2\int_{\pi/8}^{\pi} p(\phi) d\phi\\
\label{eqn:16DPSK_integral2} P_{S,2} & = & \int_{3\pi/16}^{\pi}
p(\phi)
d\phi + \int_{-5\pi/16}^{-\pi} p(\phi) d\phi\\
\label{eqn:16DPSK_integral3} P_{S,1} & = & \int_{5\pi/16}^{\pi}
p(\phi) d\phi + \int_{-11\pi/16}^{-\pi} p(\phi) d\phi
\end{eqnarray}
where $p(\phi)$ is the density of the phase angle between two
vectors perturbed by (independent) AWGN samples.  The integrals
taken in the computation of the symbol error, above, are simply the
integrals over the regions of received angle $\Delta \phi_k$ for
which an error would occur (based on the $\beta$-DSs given in
Fig.~\ref{16DPSK_decisions}).  A good approximation for integrals of
the form in (\ref{eqn:16DPSK_integral1}) to
(\ref{eqn:16DPSK_integral3}) is given in \cite{Pawula_ITC0882},
eqn.~(46), rewritten in a slightly different form below:
\begin{eqnarray} \label{eqn:Pawula}
\int_{\pi/M}^{\pi} p(\phi) d\phi \cong \frac{1}{2}
\sqrt{\frac{1+\cos(\pi/M)}{2\cos(\pi/M)}}
\textrm{erfc}\sqrt{\gamma(1-\cos(\pi/M))}, ~~~M\geq3,
\end{eqnarray}
where $\textrm{erfc}(x)\triangleq
\frac{2}{\sqrt{\pi}}\int_x^{\infty}e^{-t^2}dt$ is the complementary
error function.

\begin{figure} [t]
  \centering
  \includegraphics [width=4.5in] {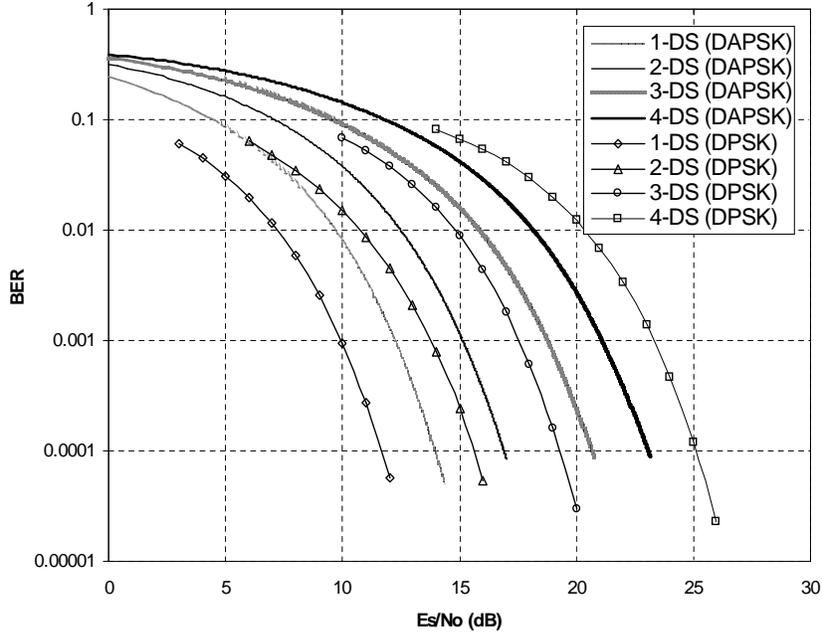}
  \caption{BER for recovering 1, 2, 3, and 4 most likely bits for 16-DPSK and 16-DAPSK (with $R=2$).}
  \label{16DAPSK_and_16DPSK_BER}
\end{figure}

Using (\ref{eqn:16DPSK_integral1}) to (\ref{eqn:Pawula}), and making
the approximation that the BER for a $\beta$-DS is $\frac{1}{\beta}
P_{S,\beta}$, the probability of error curves for 16-DPSK are given
in Fig.~\ref{16DAPSK_and_16DPSK_BER}.  The figure also provides the
BER curves (from Fig.~\ref{16DAPSK_estimate_vs_exact}) for 16-DAPSK
with $R=2$.  Note the initially surprising result that although
16-DAPSK outperforms 16-DPSK when recovering all four bits (as is
well known), when recovering 1, 2, or 3 bits 16-DPSK shows a clear
advantage over 16-DAPSK.

The fact that 16-DPSK outperforms 16-DAPSK for lower DS can be
explained by noting that in a 16-DPSK scheme, certain bits are very
well protected while others are more susceptible to error.  By
erasing the error prone bits (for a given received angle) large
gains are observed; however, for 16-DAPSK the bits are more
uniformly reliable/unreliable with the consequence that erasing the
least reliable among them will not result in gains as significant as
for 16-DPSK.  More specifically, for 16-DPSK, the least reliable bit
(in general) is the bit that changes value between two differential
angles denoting two ``adjacent" symbols.  When this bit is dropped
(e.g., for 3-DS), the performance of the remaining bits is
reminiscent of an 8-DPSK system---i.e., a significant gain is
observed.  With 16-DAPSK, an advantage is gained for 4-DS since the
minimum differential angle separating differential symbols is
$\pi/4$ (as opposed to $\pi/8$ for 16-DPSK).  Of course, this
increase in minimum angle is achieved by allowing for two distinct
amplitudes: $A_1$ and $A_2$.  This strategy does indeed succeed in
reducing the error rate of the least reliable bit.  Often, however,
the least reliable bit for a 16-DAPSK constellation is in fact the
differential amplitude bit; thus, when this is neglected, the
remaining bits are less reliable than for a 16-DPSK 3-DS since the
inner ring necessarily has a smaller radius than the standard
16-DPSK radius (assuming equal symbol energies for the two
constellations).

\begin{figure} [t]
  \centering
  \includegraphics [width=4.5in] {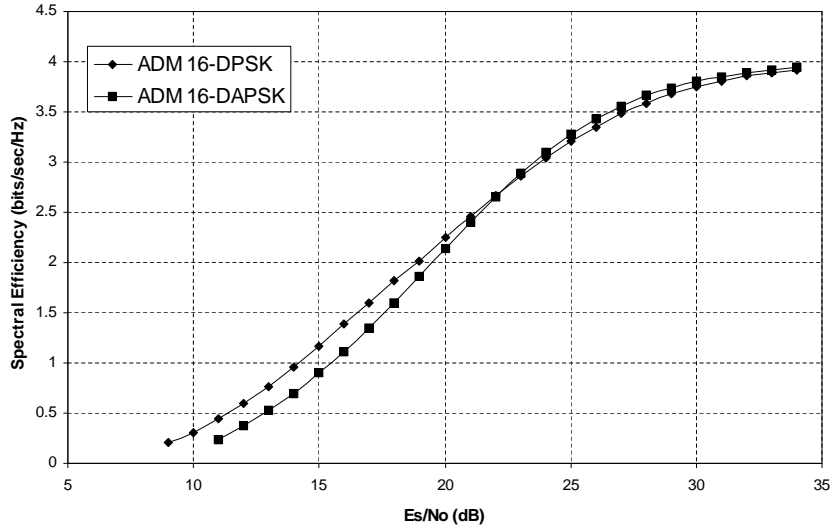}
  \caption{Spectral Efficiency for 16-DPSK and 16-DAPSK (with $R=2$).}
  \label{16DAPSK_and_16DPSK_SpecEff}
\end{figure}

With the BER computed for all $\beta$-DSs, it is now possible to
compute the optimal operating regions for the ADM system over a
Rayleigh fading channel (i.e., compute which $\beta$-DS to use for a
given observed instantaneous SNR). These operating regions are
computed using the techniques outlined in \cite{David_QBSC2006}
assuming that the fading is constant over at least two symbol
intervals. Using the computed operating regions, the spectral
efficiency (for ``raw" uncoded bits) for ADM using 16-DPSK and
16-DAPSK is given in Fig.~\ref{16DAPSK_and_16DPSK_SpecEff}. It is
seen that ADM using 16-DPSK outperforms ADM using 16-DAPSK for low
rate transmission (until about 2.5 bits per symbol); additionally,
it does not suffer a large performance deficit at higher rates. This
is explained by the relatively poorer performance the 1-DS, 2-DS,
and 3-DS of 16-DAPSK compared to 16-DPSK as noted in
Fig.~\ref{16DAPSK_and_16DPSK_BER}.

\begin{figure} [t]
  \centering
  \includegraphics [width=4.6in] {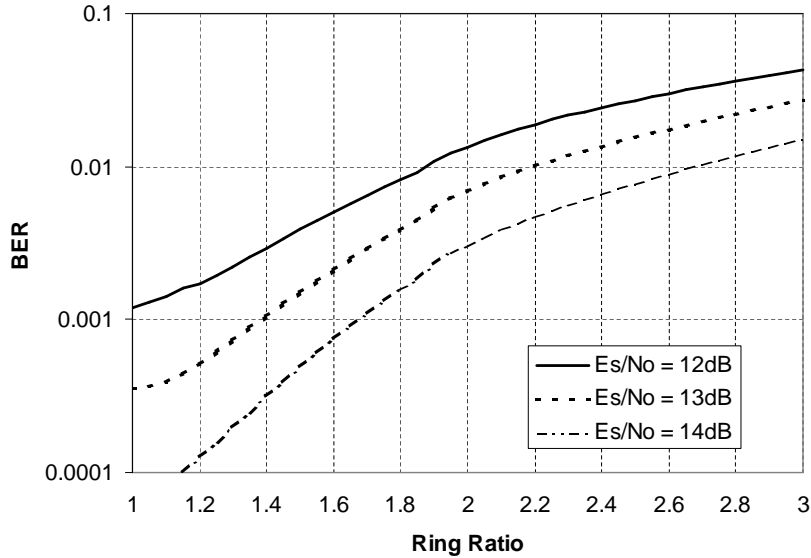}
  \caption{BER of 16-DAPSK $2$-DS for various ring ratios.}
  \label{Ring_ratios}
\end{figure}

Although \cite{Chow_EL0892} demonstrated that a ring ratio of
$R\approx2$ is an optimal choice for 16-DAPSK operating in a
Rayleigh fading environment, it is not obvious that this ring ratio
remains optimal for an ADM system using 16-DAPSK. Reference
\cite{David_Thesis2008} demonstrates the effect of the ring ratio on
the BER performance (in AWGN) for each $\beta$ decision scheme. It
is shown in \cite{David_Thesis2008} that for the 4-DS (i.e.,
standard 16-DAPSK differentially coherent demodulation), the BER is
minimized for a ring ratio of $R=2$, a result that agrees with
\cite{Chow_EL0892}. However, for the 3-DS, 2-DS, and 1-DS, it
observed that the BER tends to increase with $R$, exhibiting no
local minimum at $R=2$ (e.g., see Fig.~\ref{Ring_ratios} for 2-DS).
Thus, for all $\beta$-DS except 4-DS, the BER performance improves
for lower ring ratios. These results can be explained by observing
that as $R$ decreases below 2, the inner and outer ring radii become
closer in value; consequently $b_0$ becomes increasingly unreliable.
This results in an increase in BER for 4-DS.  For all other
$\beta$-DS, however, the unreliable $b_0$ is most likely the bit
that will be discarded; the remaining bits become more reliable for
smaller $R$ since the radius of the inner ring increases, leading to
more reliable angle measurements.

Since the BER of all $\beta$-DS (including 4-DS) increases when
$R>2$, we conclude that it is never advantageous (at least in terms
of BER performance) to use ring ratios larger than 2 in the ADM
system.  The question still remains, however, what is the best ring
ratio ($R \leq 2$) for ADM.  The answer depends on the desired
system performance and range of operation.  For example, if the
system is expected to operate primarily in high average SNR at close
to 4 bits per symbol, then $R=2$ is a good choice since the system
will spend most of its time using $4-DS$.  However, if the system is
expected to have a very broad range of operation, a smaller $R$ may
be advisable.

\begin{figure} [t]
  \centering
  \includegraphics [width=4.5in] {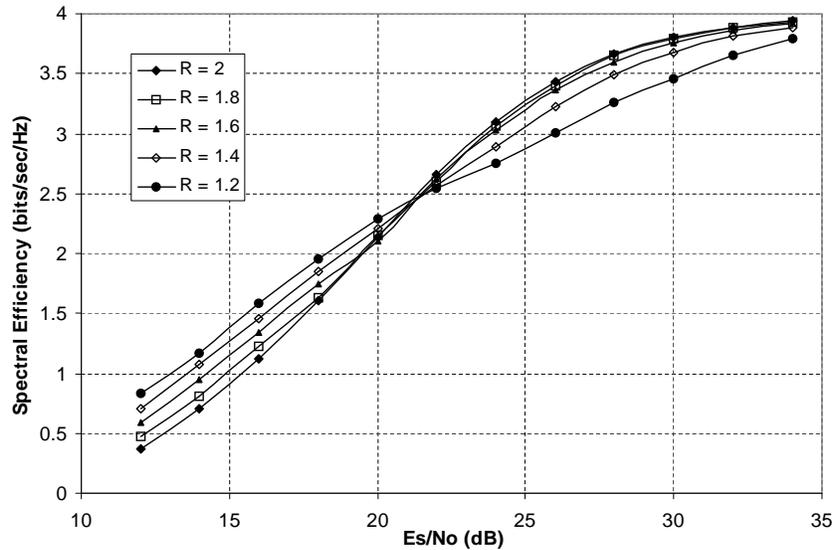}
  \caption{Spectral Efficiency of 16-DAPSK for various ring ratios (raw BER of $10^{-4}$).}
  \label{Spectral_Efficiency_RR}
\end{figure}

In Fig.~\ref{Spectral_Efficiency_RR}, the spectral efficiency curves
(for Rayleigh fading) are computed for a range of ring ratios.  As
expected, the ADM systems using low values of $R$ outperform the ADM
system using $R=2$ at lower rates; however, these systems are less
effective at higher rates when the performance of the 4-DS becomes
more significant.

\section{Conclusion}\label{conclusion}
In this paper, we have derived optimum and near-optimum receivers
for differentially coherent reception in Adaptive Demodulation
systems using 16-DPSK and 16-DAPSK. Simple decision rules were
derived for the optimal demodulation of the $\beta$ most likely bits
for both 16-DPSK and 16-DAPSK. Although these simple rules led to
simple decision schemes for 16-DPSK, the resulting decision schemes
for 16-DAPSK contained nonlinear boundaries that were not conducive
to a simple implementation.  Based on the 16-DAPSK decision rules,
heuristic near-optimal decision schemes were proposed for 16-DAPSK
with a simple implementation; in addition, formulas were derived to
tailor the decision schemes for any desired ring ratio. The
probability of error for 16-DPSK and 16-DAPSK receivers demodulating
the most reliable 1, 2, 3, and 4 bits per symbol was computed along
with the spectral efficiency of the ADM systems. A surprising result
was demonstrated that over a large operating region, 16-DPSK
actually outperforms 16-DAPSK for Adaptive Demodulation systems. The
impact of the ring ratio on the spectral efficiency of ADM using
16-DAPSK was investigated and it was shown that ring ratios larger
than 2 are never beneficial (in terms of spectral efficiency) for
such systems. A tradeoff was shown to exist where low ring ratios
improved the performance of 16-DAPSK for low average SNR operating
regions, while larger ring ratios led to an improvement in the
performance of high average SNR operating regions.


\begin{thebibliography}{10}

\bibitem{David_PIMRC2007}
J.~D. Brown, K.~N. Plataniotis, and S.~Pasupathy,
\newblock ``Adaptive demodulation with differentially coherent detection,''
\newblock in {\em Proc. of IEEE Int. Symposium on Personal, Indoor and Mobile
  Radio Communications (PIMRC'07)}, 2007.

\bibitem{Blogh_Book2002}
J.~S. Blogh and L.~Hanzo,
\newblock {\em Third-Generation Systems and Intelligent Wireless Networking:
  Smart Antennas and Adaptive Modulation},
\newblock John Wiley and Sons, 2002.

\bibitem{NandaICM0100}
S.~Nanda, K.~Balachandran, and S.~Kumar,
\newblock ``Adaptation techniques in wireless packet data services,''
\newblock {\em IEEE Communications Magazine}, vol. 38, no. 1, pp. 54--64, Jan.
  2000.

\bibitem{Zhang_ICCT2003}
P.~Zhang and L.~Li,
\newblock ``Research on beyond 3{G} mobile communications,''
\newblock in {\em Proc. of International Conference on Communications
  Technology}, April 2003, vol.~1, pp. 28--31.

\bibitem{Goldsmith_ITC1097}
A.~J. Goldsmith and S.-G. Chua,
\newblock ``Variable-rate variable-power {MQAM} for fading channels,''
\newblock {\em IEEE Trans. on Commun.}, vol. 45, no. 10, pp. 1218--1230, Oct.
  1997.

\bibitem{Metzner_ITC0684}
J.~J. Metzner,
\newblock ``An improved broadcast retransmission protocol,''
\newblock {\em IEEE Trans. on Commun.}, vol. 32, no. 6, pp. 679--683, June
  1984.

\bibitem{DavidITC0906}
J.~D. Brown, S.~Pasupathy, and K.~N. Plataniotis,
\newblock ``Adaptive demodulation using rateless erasure codes,''
\newblock {\em IEEE Trans. on Commun.}, vol. 54, no. 9, pp. 1574--1585, Sept.
  2006.

\bibitem{David_Thesis2008}
J.~D. Brown,
\newblock {\em Adaptive Demodulation Using Rateless Erasure Codes},
\newblock Ph.D. Thesis, University of Toronto, 2008.

\bibitem{LubyFOCS2002}
M.~Luby,
\newblock ``{LT} codes,''
\newblock in {\em Proc. of the 43rd Annual IEEE Symposium on Foundations of
  Computer Science (FOCS)}, 2002, pp. 271--280.

\bibitem{Shokrollahi_ITIT0606}
A.~Shokrollahi,
\newblock ``Raptor codes,''
\newblock {\em IEEE Trans. on Inform. Theory}, vol. 52, no. 6, pp. 2551--2567,
  June 2006.

\bibitem{Liu_ITWC0609}
X.~Liu and T.~J. Lim,
\newblock ``Fountain codes over fading relay channels,''
\newblock {\em IEEE Trans. on Wireless Commun.}, vol. 8, no. 6, pp. 3278--3287,
  June 2009.

\bibitem{Illanko_VTC2009}
K.~Illanko and A.~Anpalagan,
\newblock ``Cooperative communication using bit-selective adaptive demodulation
  and {R}aptor codes: The {G}aussian relay channel case,''
\newblock in {\em Proc. of IEEE 69th Vehicular Technology Conference, VTC
  Spring}, 2009.

\bibitem{Stojanovic_JSAC0905}
M.~Stojanovic,
\newblock ``An adaptive algorithm for differentially coherent detection in the
  presence of intersymbol interference,''
\newblock {\em IEEE Journal on Selected Areas in Commun.}, vol. 23, no. 9, pp.
  1884--1890, Sept. 2005.

\bibitem{Lee_IT0781}
J.~S. Lee, R.~H. French, and Y.~K. Hong,
\newblock ``Error performance of differentially coherent detection of binary
  {DPSK} data transmission on the hard-limiting satellite channel,''
\newblock {\em IEEE Trans. on Inform. Theory}, vol. IT-27, no. 4, pp. 489--497,
  July 1981.

\bibitem{Kahn2006}
J.~M. Kahn,
\newblock ``Modulation and detection techniques for optical communication
  systems,''
\newblock in {\em Proc. of Optical Amplifiers and Their Applications/Coherent
  Optical Technologies and Applications}. Technical Digest OAA/COTA06, paper
  CThC1, 2006.

\bibitem{Proakis2001}
J.~G. Proakis,
\newblock {\em Digital Communications},
\newblock New York: McGraw-Hill, forth edition, 2001.

\bibitem{Jamshid_BSC2010}
J.~Abouei, J.~D. Brown, K.~N. Plataniotis, and S.~Pasupathy,
\newblock ``On the energy efficiency of {LT} codes in proactive wireless sensor
  netwroks,''
\newblock {\em Proc. of IEEE Biennial Symposium on Communication (QBSC'10),
  Kingston, Canada}, pp. 114--117, May 2010.

\bibitem{KschischangITIT0201}
F.~R. Kschischang, B.~J. Frey, and H.-A. Loeliger,
\newblock ``Factor graphs and the sum-product algorithm,''
\newblock {\em IEEE Trans. on Inform. Theory}, vol. 47, no. 2, pp. 498--519,
  Feb. 2001.

\bibitem{Lindsey_Book1973}
W.~C. Lindsey and M.~K. Simon,
\newblock {\em Telecommunication Systems Engineering},
\newblock Englewood Cliffs, NJ: Prentice-Hall, 1973.

\bibitem{Xiao_JCN2006}
L.~Xiao, X.~Dong, and T.~T. Tjhung,
\newblock ``Maximum likelihood receivers for {DAPSK} signaling,''
\newblock {\em Journal of Communications and Networks}, vol. 8, no. 2, pp.
  205--211, June 2006.

\bibitem{Divsalar_ITC0390}
D.~Divsalar and M.~K. Simon,
\newblock ``Multiple-symbol differential detection of {MPSK},''
\newblock {\em IEEE Trans. on Commun.}, vol. 38, no. 3, pp. 300--308, March
  1990.

\bibitem{Xiao_Thesis2004}
L.~Xiao,
\newblock {\em Signalling Constellations in Wireless Fading Channels},
\newblock M.Sc. Thesis, University of Alberta, 2004.

\bibitem{Chow_EL0892}
Y.~C. Chow, A.~R. Nix, and J.~P. McGeehan,
\newblock ``Analysis of 16-{APSK} modulation in {AWGN} and {R}ayleigh fading
  channel,''
\newblock {\em Electron. Lett.}, vol. 28, pp. 1608--1610, Aug. 1992.

\bibitem{Pawula_ITC0882}
R.~F. Pawula, S.~O. Rice, and J.~H. Roberts,
\newblock ``Distribution of the phase angle between two vectors perturbed by
  {G}aussian noise,''
\newblock {\em IEEE Trans. on Commun.}, vol. 30, no. 8, pp. 1828--1841, Aug.
  1982.

\bibitem{David_QBSC2006}
J.~D. Brown, K.~N. Plataniotis, and S.~Pasupathy,
\newblock ``Adaptive demodulation performance over a {R}ayleigh fading
  channel,''
\newblock in {\em Proc. of 23rd Biennial Symposium on Communications}.
  Kingston, ON, Canada, May 2006, pp. 51--54.

\end{thebibliography}
\end{document}